\documentclass[conference]{IEEEtran}
\IEEEoverridecommandlockouts
\usepackage{cite}
\usepackage{amsmath,amssymb,amsfonts}
\usepackage{algorithmic}
\usepackage{graphicx}
\usepackage{textcomp}
\usepackage{xcolor}
\usepackage{url}
\usepackage{enumitem}
\setlist{topsep=2pt, itemsep=1pt, parsep=0pt} 
\def\BibTeX{{\rm B\kern-.05em{\sc i\kern-.025em b}\kern-.08em
    T\kern-.1667em\lower.7ex\hbox{E}\kern-.125emX}}

\usepackage{fancyhdr}
\begin{document}


\title{Me Among Us: Affective Framing in Data Donation
}

\author{\IEEEauthorblockN{Zeya Chen}
\IEEEauthorblockA{\textit{Institute of Design (ID)} \\
\textit{Illinois Institute of Technology}\\
Chicago, USA\\
zchen103@hawk.illinoistech.edu}
\and
\IEEEauthorblockN{Zach Pino}
\IEEEauthorblockA{\textit{Institute of Design (ID)} \\
\textit{Illinois Institute of Technology}\\
Chicago, USA\\
zach.pino@illinoistech.edu}
}

\maketitle
\thispagestyle{fancy}


\begin{abstract}
This study investigates how different framing approaches influence the affective aspects of data donation decision-making. Although framing effects are well studied in charitable giving, how affective framing shapes data donation, especially through data visualization, remains poorly understood. Using a theoretical framework based on the functions of affect in decision-making, we examine how three distinct framing approaches, an individual-donor lens (Group A), an individual-collective lens (Group B), and a collective-institutional lens (Group C), shape participants' affective experiences and subsequent donation decisions. Through a real-world data donation study (N=24), we found that framing designs substantially influenced donation outcomes, with the individual-collective lens generating the most favorable responses. Our analysis illustrates how affect can functions as information, motivation, and as a spotlight during the decision-making process, providing insights for designing more informed data donation interfaces and communications. This research contributes to understanding the complex interplay between framing designs, affective responses, and decision outcomes in data donation contexts.

\end{abstract}


\begin{IEEEkeywords}
affective decision-making, data donation, framing effect, data visualization
\end{IEEEkeywords}


\section{Introduction}
Data donation, the voluntary contribution of personal data primarily for common good purposes, has emerged as a promising approach for accessing rich, individual-level data to support various research and development initiatives \cite{skatova_psychology_2019}. As organizations increasingly seek access to personal data, understanding how to ethically facilitate data donation becomes crucial. Unlike traditional data collection methods, data donation is considered a donor-centered approach where individuals actively opt-in to contribute their personal data without immediate compensation\cite{gomez_ortega_beyond_2023}.

A critical challenge in data donation is creating informed consent experiences that adequately communicate complex information about data use while supporting autonomous decision-making\cite{gomez_ortega_beyond_2023}. Previous research suggests that traditional text-based consent forms often fail to provide sufficient understanding of what data is being collected and how it will be used\cite{pins_alexa_2021}. Data visualization offers a promising approach to address this challenge by making abstract data more concrete and comprehensible\cite{kennedy_engaging_2016,he_enthusiastic_2024}.

The presentation of information significantly influences decision-making, a phenomenon well-documented in behavioral science as the ``framing effect'' \cite{hoover_moral_2018}. While framing effects have been extensively studied in traditional donation contexts (e.g., charitable giving, organ donation), little is known about how different framing approaches affect data donation decisions, particularly when visualization is used as the medium for information presentation. Prior work established that this framing shapes donation behavior as a behavioral design lever \cite{chen_framing_2026}.  Yet those decisions are emotionally charged, entangling privacy, trust, and self-disclosure, suggesting that affect—not cognition alone—may be central to how framing operates.

We therefore examine how framing approaches in data exploration shape the affective aspects of data donation decision-making. Drawing on established theories of affect in decision-making, we study three framing approaches:

\begin{itemize}
    \item \textbf{Individual-donor lens} (Group A): Framing that focuses on personal insights and self-discovery
    \item \textbf{Individual-collective lens} (Group B): Framing that emphasizes social comparison and relationships
    \item \textbf{Collective-institutional lens} (Group C): Framing highlights collective contribution and institutional benefits

\end{itemize}
By examining how these framings shape participants' affective experiences and donation decisions, we contribute to affective computing in decision-support systems and offer practical insights for designing affect-aware data donation experiences.

\section{Related Work}
\subsection{Data Donation and Informed Consent}
Data donation is the process by which a person transfers their personal data, without expecting immediate return, to an entity that will use it in a specified context \cite{skatova_psychology_2019, prainsack_data_2019}. Unlike aggregate collection such as web scraping or proprietary platform APIs, it is donor-driven: individuals opt in to share their own data directly with researchers. Governed by regulations such as the GDPR right to data portability \cite{ohme_digital_2022}, it grants individuals control over what they share and has drawn interdisciplinary interest spanning privacy-preserving data management \cite{boeschoten_framework_2022}, practice guidelines \cite{carriere_best_2024}, and platform infrastructure \cite{gomez_ortega_participation_2024, hase_fulfilling_2024}.

A central challenge is obtaining truly informed consent \cite{gomez_ortega_beyond_2023}, which should cover why data is collected, how it is used and stored, and what risks exist \cite{sloan_linking_2020}. Text-based consent forms have proven inadequate: users struggle with vague policies \cite{pins_alexa_2021, oconnor_privacy_2017}, and the data stays abstract at consent time \cite{gomez_ortega_beyond_2023}, so participants often misread its implications even when it is visualized \cite{groot_kormelink_meaningful_2025}.

In response, workflows increasingly treat consent as an ongoing process, using local processing \cite{boeschoten_framework_2022}, inspection and deletion \cite{boeschoten_port_2023, pfiffner_data_2024}, and interactive interpretation \cite{franzen_communicating_2024, gomez_ortega_participation_2024}, with data exploration to make abstract data concrete for donors.

\subsection{Framing Effects and Information Presentation}
How information is presented, the ``framing effect,'' can alter decisions more than the choice itself \cite{tversky_framing_1981, fagley_effects_2010}. In donation contexts, solicitation framing shapes charitable giving beyond individual and situational differences \cite{hoover_moral_2018}, with self- versus other-beneficial frames moderating donation \cite{fielding_materialists_2020, keser_charitable_2023, sneddon_personal_2020}.

The individual-collective dichotomy is especially relevant here. Skatova and Goulding \cite{skatova_psychology_2019} find collective benefits and social duty the strongest predictors of data-donation intention, while self-serving motives predict negatively. Consistent with this, ``help others'' framing raises charitable giving over ``help self'' \cite{fielding_materialists_2020}, and collective-led frames outperform individual-led ones \cite{keser_charitable_2023}. Social norms and in-group comparison are strong prosocial levers \cite{allcott_social_2011, anderson_behavioral_2014}. Social-comparison nudges, for example, have been used to encourage healthier choices \cite{dicosola_nudging_2022}, suggesting that the individual-collective lens \cite{lindauer_comparing_2020} suits data-donation decisions.

Visualization is likewise never affectively neutral: design techniques shape interpretation \cite{hullman_visualization_2011}, and charts projecting an ``illusion of objectivity'' can mislead more than those acknowledging perspective \cite{dignazio_data_2023}. More broadly, analytical and data-source choices shape the conclusions drawn from data \cite{tu_study_2025}. Researchers thus advocate ``neutrality-adjacent goals'' \cite{lan_negative_2022}, and affective-visualization work shows visualizations influence not only understanding but emotions, values, and beliefs \cite{lee-robbins_affective_2023}, suggesting visualization framing may shape data-donation decisions through affect.

\subsection{Affect in Decision-Making}
In affective science, affect refers to low-dimensional, continuous states that vary along two dimensions: valence (positive-to-negative evaluation) and arousal (perceptual activation). While traditional decision-making research relied on rational choice models, recent works acknowledge the central role of affect in decision processes\cite{lindauer_comparing_2020, asutay_affective_2024}.

The historical dichotomy between rational and emotional decision-making has been increasingly challenged. In the donation context, while emotional appeals are often found more persuasive than rational appeals in charitable giving \cite{vieira_altruistic_2025}, other experiments find no significant difference when rational appeals are well-designed \cite{lindauer_comparing_2020}. Similarly, Kennedy and Hill\cite{kennedy_engaging_2016} critique the ``reason/emotion binary'' in visualization research, while Lan et al.\cite{lan_negative_2022} demonstrate that emotion can co-exist with and even amplify rational data interpretation.

Previous research has identified three major functions of affect in behavioral decision-making\cite{asutay_affective_2024}:

\begin{itemize}
    \item Affect-as-information: Affect assigns value to the object of judgment,  providing insights that guide decisions.
    \item Affect-as-motivation: Affect motivates decisions through goal-directed behaviors, acting as a control system that signal progress toward goal attainment.
    \item Affect-as-a-spotlight: Affect shifts attention and influences the weighting of information in decisions, focusing cognitive processing on particular aspects of decision contexts.

\end{itemize}

These affective mechanisms operate through a process called ``affective integration,'' whereby signals important for decision-making prompt changes in the individual's affective state, which are integrated over time into a unified overall affective experience\cite{asutay_affective_2024}. This dynamic integration is shaped by context and environmental cues \cite{liu_examining_2023}, suggesting that framing approaches might significantly influence affective integration in data donation decisions.

\section{Method}

To examine these framing effects in practice, we studied data donation using students' institutional calendar data at the Institute of Design (ID) at Illinois Institute of Technology. ID's administration already has exclusive access to this data, and many students are unaware that it does. Following Carriere et al.'s best-practice guideline for data donation studies \cite{carriere_best_2024}, we chose calendar data because it is valuable to ID for understanding scheduling patterns, resource allocation, and community engagement, while also containing personal information. 
Because students often use their institutional accounts for personal life as well, the data can reveal not only academic commitments and collaborations but also personal appointments, relationships, and daily routines, giving it a meaningful degree of sensitivity.

The study offered participants a genuine opportunity to donate their anonymized calendar data to ID for research purposes. This real-world context was essential for observing authentic rather than hypothetical decision-making. It also defined our target participants: ID's currently enrolled graduate students. A companion paper analyzes the same study through a behavioral design lens \cite{chen_framing_2026}, whereas the present paper focuses on the affective mechanisms underlying the donation decision.

\subsection{Exploration Framing Design}
We designed three distinct framing approaches to investigate how different conceptual lenses influence data donation decision-making (see the top table of Fig. \ref{fig:framing}):

\begin{figure*}[t]
    \centering
    \includegraphics[width=\textwidth]{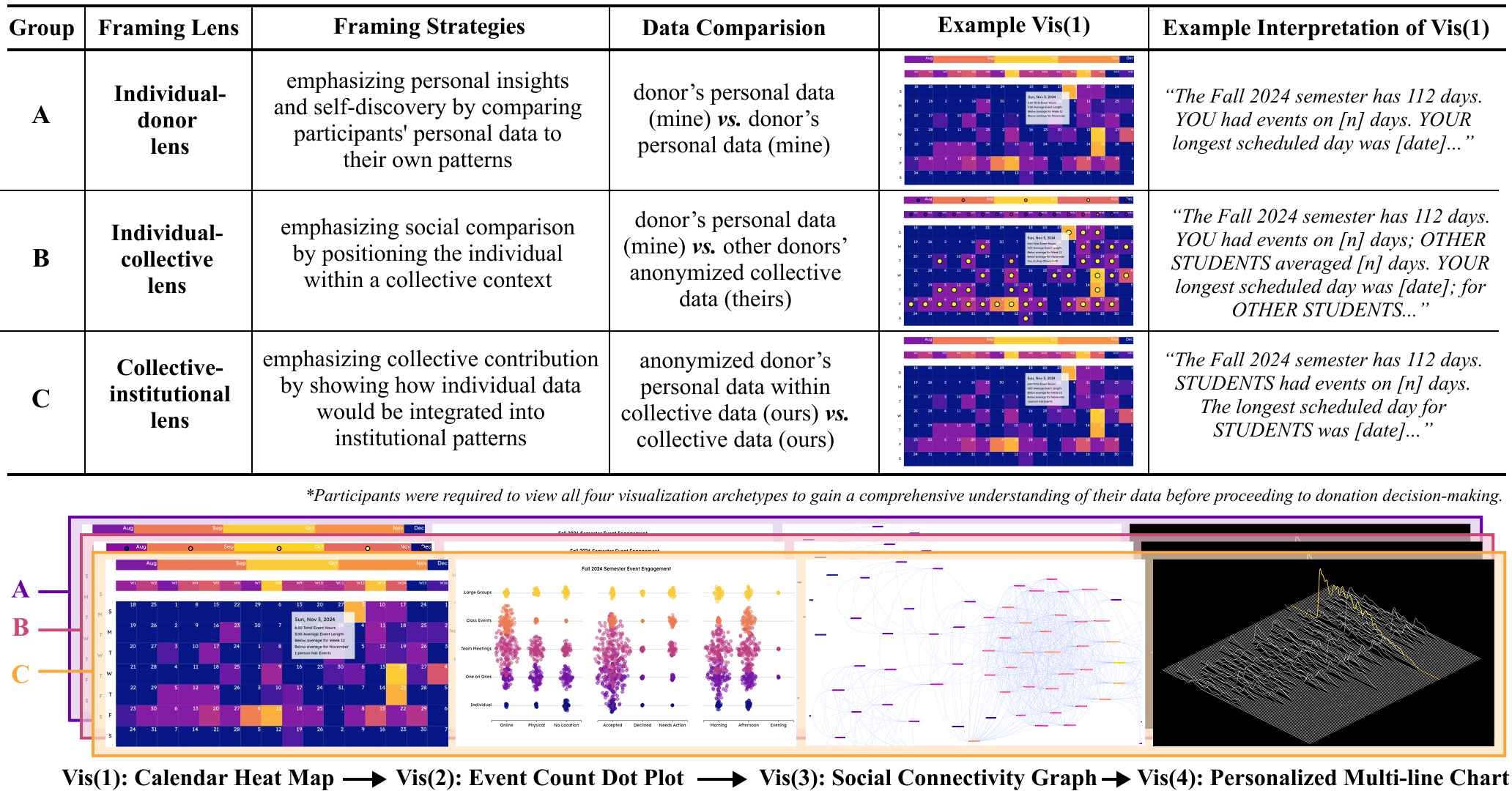}
    \caption{Framing conditions (top) and the four visualization archetypes viewed by all groups (bottom).}
    \label{fig:framing}
\end{figure*}

\begin{itemize}
    \item \textbf{Group A – Individual-donor lens}: This framing emphasized personal insights and self-discovery by highlighting patterns in participants' personal data.
    \item \textbf{Group B – Individual-collective lens}: This framing emphasized social comparison by positioning the individual within a collective context.
    \item \textbf{Group C – Collective-institutional lens}: This framing emphasized collective contribution by showing how individual data would be integrated into institutional patterns.
\end{itemize}

The complete per-condition framing text and annotated visualizations are provided in the Supplement, S5.

\subsection{Data Visualization Design}
All visualizations followed Sloan et al.\cite{sloan_linking_2020}'s data donation consent template and Ortega et al.\cite{gomez_ortega_beyond_2023}'s data informed framework to include critical data information (goals, types, amount, processing, etc.). We implemented a neutrality-adjacent approach \cite{lan_negative_2022} and rhetoric visualization framework \cite{hullman_visualization_2011} to present data accurately while acknowledging inevitable framing choices. Four interactive visualizations were presented in sequence (see the bottom of Fig. \ref{fig:framing}):
\begin{enumerate}
    \item Calendar Heat Map: A temporal visualization showing event frequency and duration across days and hours, allowing participants to explore patterns over time.
    \item Event Count Dot Plots: Category-based visualizations displaying distributions of meeting types, locations (online/in-person), and other calendar event attributes.
    \item Social Connectivity Graph: A network visualization revealing connections between calendar events and attendees, highlighting social dimensions of calendar data.
    \item Personalized Multi-line Chart: An interactive visualization allowing participants to select variables of interest and create aesthetic data art representing their calendar data.
    \end{enumerate}

All visualizations offered accessibility options (reduced motion to limit animations, a high-contrast mode, and an alternative palette designed for color vision differences) and used the Lexend typeface~\cite{shaver-troup_lexend_2018} throughout for legibility. All calendar data was processed locally after download to preserve privacy.

\subsection{Participants and Procedure}
We initially recruited 30 graduate students who regularly used their institutional Google Calendar. After removing pilot participants (n=3) and incomplete participations (n=3), our final sample consisted of 24 participants equally distributed across the three framing groups (8 per group). Participants were 18-40 years old (M=27.0), with 18 female and 6 male participants, representing different academic levels and institutional data knowledge. All participants were enrolled at ID, a graduate-only design school with a majority-female student body, which is reflected in our sample's exclusively Master's and PhD composition, and its greater share of women.

The study followed a structured procedure:
\begin{enumerate}
    \item Consent and pre-study survey: Participants affirmed informed consent and completed a survey about their calendar usage, privacy attitudes, and demographic information.
    \item Data exploration: Participants granted temporary access to their calendar data, which was processed locally to generate the visualizations. They then explored their data through the four visualization archetypes, with framing determined by their assigned condition. During exploration, participants provided think-aloud~\cite{ericsson_protocol_1993} comments (prompts in the Supplement, S3) and ratings of their understanding.
    \item Donation decision-making: After exploration, participants decided whether to donate their anonymized calendar data.
    \item Post-study survey and  interview: Participants completed a survey with open questions about their experience and 7-point Likert scale questions, measuring their feelings toward their donation decision, perceived helpfulness, understanding of data characteristics, potential risks, perceived benefits, and other factors.
\end{enumerate}

All participants volunteered to join the study, and were not compensated for their time. Because participants were graduate students at the researchers' own institution, recruitment and consent materials stated, in writing and verbally, that the decision to donate would not affect their academic standing or relationship with the institution.

\subsection{Data Collection and Analysis}
We collected multiple forms of data through behavioral, cognitive, and affective measurement \cite{sukei_predicting_2021} to comprehensively capture participants' experiences. Following established protocols in affective science \cite{li_scale_2013}\cite{luong_valuing_2023}, we used 7-point Likert scales to measure affective valence and intensity. Each item was rated from 1 (strongly disagree) to 7 (strongly agree); throughout, we interpret ratings of 1--3 as negative, 4 as neutral, and 5--7 as positive. Our multi-modal approach included behavioral metrics (time spent on data exploration, donation decisions), perceptual data (pre/post survey Likert scale ratings of cognitive perceptions and affective feelings), verbal reports (think-aloud protocols during data exploration~\cite{ericsson_protocol_1993}), and open-ended survey responses.


Analysis followed a mixed-methods approach. We assessed affect-as-motivation through exploration duration and perceived helpfulness, and affect-as-spotlight through pre/post changes in perceived understanding (survey items in the Supplement, S2), using think-aloud quotes to illustrate both. For affect-as-information, we applied reflexive thematic analysis \cite{braun_reflecting_2019}: a single analyst open-coded participants' emotion expressions from the think-aloud and open-ended responses and grouped them into affective patterns. Because these patterns came from one coder, inter-rater reliability does not apply as a quality criterion, and we report the findings as exploratory and illustrative (codebook in the Supplement, S4).

\section{Results}

\subsection{\textbf{Donation Decision Outcomes}}
\begin{figure*}[!t]
    \centering
    \includegraphics[width=\textwidth]{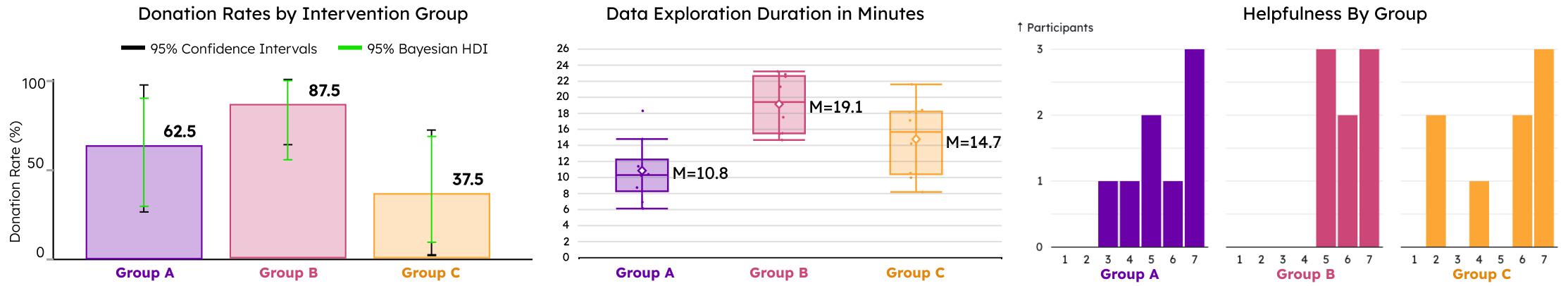}
    \caption{Quantitative outcomes by framing group. Group B (individual-collective) had the highest donation rate, longest exploration duration, and most positive helpfulness ratings, whereas Group C (collective-institutional) was the lowest and most polarized.}
    \label{fig:motivation}
\end{figure*}

We first examined donation decisions across framing conditions (Fig.~\ref{fig:motivation}, left). Of the 24 participants, 15 (62.5\%) chose to donate their data while 9 (37.5\%) declined. Donation rates varied considerably by framing lens: Group A (individual-donor lens) 62.5\%, Group B (individual-collective lens) 87.5\%, and Group C (collective-institutional lens) 37.5\%.

These outcomes suggest that framing approach meaningfully influenced participants' donation decisions. Notably, the 50-percentage-point gap between Group B and Group C represents a substantial practical difference (Cramér's V=0.42, indicating a medium-to-large effect size). The omnibus association was $\chi$²(2)=4.27, p=.12, and Fisher's exact test for the Group B versus Group C contrast gave p=.12 (Fisher's exact because several expected cell counts fell below 5). At N=24 this did not reach conventional significance; a sensitivity analysis suggests roughly 54 participants would be needed to detect an effect of this magnitude at 80\% power, so the study is powered to estimate rather than confirm. Because participants faced a genuine donation decision involving their own institutional data rather than a hypothetical scenario, each outcome reflects a real, consequential choice, lending the results an ecological validity that a larger hypothetical sample would lack. The sample represents 30–38\% of the eligible institutional population (approximately 80–100 individuals), and Bayesian estimation placed the posterior probability that Group B exceeded Group C at$\approx$0.98 (95\% HDI on the difference [0.00, 0.74]; Fig.~\ref{fig:motivation}, left). Per-condition demographics (see Supplement, S1) indicate broadly comparable groups in gender, academic level, background, and prior data-donation experience; the highest-donating group (Group B) was among the least experienced, so data literacy does not account for the effect. To understand the affective mechanisms behind these outcomes, we analyzed participants' experiences through the three functions of affect in decision-making \cite{asutay_affective_2024}: affect-as-motivation, affect-as-spotlight, and affect-as-information.

\subsection{\textbf{Affect-as-Motivation: Exploration Duration and Perceived Helpfulness}}
Affect can motivate decisions through goal-directed behavior, acting as a control system that signals progress toward goal attainment. We assessed this mechanism by examining participants' time spent on exploration and perceived helpfulness ratings.

\subsubsection{Exploration Duration}


Overall, participants spent an average of M=14.9 minutes (SD=5.30) exploring their data visualizations, ranging from 6.1 to 23.2 minutes. Exploration time differed across framing conditions, one-way ANOVA F(2, 21)=7.75, p=.003, $\eta$²=0.43 (a large effect) (Fig.~\ref{fig:motivation}, center).

\begin{itemize}
    \item Group A participants (individual-donor lens) showed moderate engagement with the least dispersion (M=10.8 minutes, SD=4.04), with participants primarily focusing on personal patterns.
    \item Group B participants (individual-collective lens) spent the longest time exploring data (M=19.1 minutes, SD=3.71); the Group A–B difference was large (Cohen's d=2.13, mean difference 8.3 minutes, 95\% CI [4.7, 11.8]), suggesting sustained engagement with the social-comparison framing. As one Group B participant explained: ``It's interesting to see comparison, such a great view, especially the overview of how many days I'm busy this semester compared to others'' (p9).
    \item Group C participants (collective-institutional lens) showed the greatest variance in engagement time (M=14.7 minutes, SD=4.80, the largest SD of the three groups), with a bimodal distribution: participants clustered into either shorter or extended (20+ minute) explorations. This polarized engagement was reflected in their comments. Some participants expressed difficulty locating their individual data within the collective lens: ``I am trying to find myself (in the social connectivity graph)...'' (p22), while others sought to understand the meaning of specific anonymized data points: ``This point represents this [group of people] I had a meeting with on a Tuesday, right?'' (p24).
\end{itemize}

No clear pattern emerged between donation decision outcome and the time spent on exploration when analyzed across all groups, suggesting that engagement time alone did not determine donation decisions.

\subsubsection{Perceived Helpfulness}
We measured perceived helpfulness of data exploration through the after-study survey question ``(After data exploration,) I know enough to make my decision about whether to donate'' (see After-Study Q5, Supplement S2), rated on a 7-point Likert scale. Results showed differences by both framing condition and decision outcome (Fig.~\ref{fig:motivation}, right). Perceived helpfulness was highest under the individual-collective lens: Group A M=5.50 (SD=1.51), Group B M=6.00 (SD=0.93), Group C M=5.12 (SD=2.17) on the 7-point scale. The association between framing and helpfulness corresponded to a medium effect (Cramér's V=0.38); given the small per-cell counts, we report it descriptively.

In general, all 15 participants who donated reported positive perceptions of data exploration helpfulness (ratings 5–7). Among the 9 non-donors, 4 reported positive perceptions, 2 were neutral, and 3 reported negative perceptions.

\begin{itemize}
    \item     Group A (individual-donor): 6 of 8 participants (75\%) reported positive perceptions, with moderate agreement (ratings 5–6) being most common (n=3).
    \item Group B (individual-collective): All 8 participants (100\%) reported positive perceptions of helpfulness (ratings 5–7), with most (n=5) indicating moderate agreement (rating 5-6).
    \item Group C (collective-institutional): 5 of 8 participants (62.5\%) reported positive perceptions, but with the highest polarization. 3 participants strongly agreed (rating 7) while 2 participants disagreed (ratings 1–3).
\end{itemize}

The qualitative data provided context for these ratings. Group B participants (p16) expressed sentiments about personal connection during exploration, such as being able to ``relate to them.'' In contrast, Group C participants (p19) expressed disappointment with the collective lens because they could not locate their individual data, questioning ``... so am I the target audience (for this data exploration)?''

\subsection{\textbf{Affect-as-Spotlight: Attention and Information Weighting}}
Affect can shift attention and influence the weighting of information in decisions. We examined this mechanism through changes in participants' perceived understanding of key donation topics before and after data exploration. Across nine consent topics rated on 7-point Likert scales, the most significant positive changes occurred for:
\begin{itemize}
    \item ``How researchers could process the donated data'' (min=-3, max=5, M=2.00, SD=1.89)
    \item ``What my institute could 'see' from my calendar data'' (min=-3, max=6, M=1.67, SD=2.10)
    \item ``How much and what types of data would be collected'' (min=-2, max=6, M=2.21, SD=2.41)
\end{itemize}
Interestingly, one topic showed a slight negative change:
\begin{itemize}
    \item ``Who has access to my institute calendar data'' (min=-5, max=4, M=-0.21, SD=2.30)
\end{itemize}

This suggests that data exploration raised new questions or concerns about data access that were not apparent from initial consent information provided, shifting participants' attention to this aspect of the donation.

Analysis by framing lens revealed that different frames directed attention to different aspects of the donation decision:

\begin{itemize}
    \item     Group A (individual-donor) showed the most positive changes in ``understanding how much and what types of data would be collected'' (min=-2, max=6, M=2.25, SD=2.60). This aligns with their qualitative comments focused on personal data categorization: ``It's interesting to look back at me, and remember what happened that day'' (p5).
    \item Group B (individual-collective) showed the strongest overall changes in perceived understanding compared to the other groups. They reported a notably high positive change in ``understanding what my institute could 'see' from my calendar data'' (min=-1, max=6, M=2.75, SD=2.43). As one participant noted: ``That's nice to see that comparison with mine versus the rest of the people... It gives me a sense of, you know, I can relate to them'' (p11). This suggests that the individual-collective framing shifted attention toward the social dynamics of the data.
    \item Group C (collective-institutional), similar to Group A, showed the largest change in ``understanding how much and what types of data would be collected''  (min=-1, max=5, M=1.75, SD=2.49) but overall showed the least change in understanding. Several participants reported negative affect such as confusion and disappointment during exploration. 
\end{itemize}

\subsection{\textbf{Affect-as-Information: Affective Responses in Decision-Making}}
Affect can assign value to the object of judgment, providing informational input to the decision process. We distinguish between integral affect (experienced while considering the object of judgment) and incidental affect (independent of the object but potentially misattributed to it) \cite{asutay_affective_2024}. 
We analyzed participants' affective responses through their think-aloud comments during exploration (in-the-moment affect) and their open-ended post-study responses and after-decision feeling ratings (retrospective affect; see After-Study Q1, Supplement S2).

Thematic analysis revealed distinct affective patterns associated with donation decisions:
    \begin{itemize}
        \item Donors typically expressed curiosity about visualizations, excitement about discovery, comfort with information sensitivity, and appreciation of benefits. For example, one donor noted, ``This is beautiful, like mountain wave, something like texture to the data'' (p5), while another recognized personal value: ``If Google Calendar had this... maybe it could help me fix my life'' (p8).
        \item Non-donors more frequently expressed confusion, lack of personal relevance, skepticism about data accuracy, and privacy concerns about exposing personal information.
    \end{itemize}

Post-decision feeling ratings revealed that all donors reported positive feelings about their decision (ratings of 5–7), while non-donors reported more mixed or uncertain feelings.

\begin{itemize}
    \item Group A (individual-donor lens) participants primarily expressed emotions related to self-discovery and personal insights. Their affective responses centered on seeing personal patterns visualized: ``It is interesting to look back at me and remember what happened that day'' (p3). Although all Group A participants (100\%) reported positive feelings in their decisions, these affective responses were less directly related to donation considerations, indicating a potential incidental affect bias in their decision-making.
    \item Group B (individual-collective lens) participants expressed emotions centered on social comparison and relationships, highlighting connection and social positioning within the group context.
    \item Group C (collective-institutional lens) participants expressed the most mixed feelings (25\% with ratings below 5) with an institutional perspective, often struggling to identify their individual position within the broader patterns or expressing preference for individual comparisons.
\end{itemize}

\section{Discussion}

\subsection{Affective Framing Effect in Data Donation}
Our study found substantial differences in donation rates across the three framing conditions (Group B: 87.5\%, Group A: 62.5\%, Group C: 37.5\%), reflecting how affective framing shaped donation decisions. Examining each group's affective experience, we trace how each framing engaged a distinct affective mechanism and how that mechanism relates to the group's donation outcome.

\textit{Group A (individual-donor lens):} participants showed moderate exploration and mostly positive helpfulness (75\%), but their affect centered on personal discovery and self-reflection, often disconnected from the donation. This ``incidental affect,'' together with a negative shift in understanding ``who has access'' to their data, left their emotions misaligned with the sharing decision and may help explain their intermediate donation rate (62.5\%).

\textit{Group B (individual-collective lens):} participants in this group showed the most consistent positive engagement, sustained exploration, and unanimously positive helpfulness ratings. By positioning personal data within a social context, this framing produced the strongest gains in perceived understanding, especially of institutional visibility, and elicited affect directly relevant to the donation decision such as expressions of social connection and contextual understanding. This decision-relevant "integral affect" aligned engagement, understanding, and emotion, which may help explain the group's high donation rate.

\textit{Group C (collective-institutional lens):} participants showed polarized engagement and the most divided helpfulness. Their affect often centered on difficulty locating themselves within the collective view, with effortful attempts to find their own data. This group showed the smallest understanding gains, along with emotions of confusion or disengagement rather than decision-relevant feelings, consistent with their lower rate (37.5\%).

Across mechanisms, affective framing shaped outcomes by motivating engagement, directing attention, and generating decision-relevant emotions. Group B showed the highest donation rate, alongside the strongest alignment between participants' affective states and the decision context.

These patterns tie back to the three strands in Section II. First, they qualify the individual-collective framing in donation research \cite{fielding_materialists_2020, keser_charitable_2023, skatova_psychology_2019}: collective framing without a personal anchor (Group C) lowered donation, so the effect reflects the individual-collective blend (Group B) rather than collective appeals alone. Second, they extend the emotion-versus-reason account of donation \cite{lindauer_comparing_2020, kennedy_engaging_2016}: decision-relevant integral affect (Group B) supported informed choice, whereas Group A's incidental affect did not. Third, they reinforce that visualization framing is never neutral but inherently affective \cite{hullman_visualization_2011, dignazio_data_2023, lee-robbins_affective_2023}.

\subsection{Affective Framing as a Design Strategy}
Framing approaches shape the affective mechanisms of data donation decisions, pointing to a design perspective for affective computing. Whereas affective computing often detects, responds to, or simulates emotions through interface elements, our work highlights conceptual framing as a lever for shaping affective experience and subsequent decisions.

The individual-collective lens (Group B) had the highest donation rate (87.5\%); this reflected how the whole experience was conceptually framed to balance personal relevance with collective benefit, not only in its interface or visualization designs. Accordingly, affective computing should expand beyond emotional interface design and interaction-level techniques \cite{yu_aerorigui_2023} to include what we term ``affective framing'' as a fundamental design strategy.

Affective framing influences how users interpret and emotionally respond to the entire interaction experience. For HCI designers working with emotionally complex contexts like data privacy and donation \cite{chen_speculating_2025, chen_privacymotiv_2026}, this approach offers a new intervention point that does not require emotion detection technologies but instead leverages conceptual framing to shape the affective journey through critical decision points, complementing behavioral strategies such as ``positive friction'' that support deliberate decision-making \cite{chen_positive_2024}.

\subsection{Temporal Dynamics of Affect in Decision Processes}
Our study reveals that affective mechanisms are not static but shift in prominence across the decision journey. Affect-as-motivation was strongest during initial engagement, where Group B's framing sustained exploration and Group C's framing produced polarized responses. During data interpretation, the spotlight dimension became more prominent: Group B participants focused on social patterns while Group A remained anchored in personal insights. In the pre-decision phase, affect-as-information dominated, with Group A relying on incidental affect (excitement about personal discoveries) and Group B integrating decision-relevant integral affect. This temporal perspective suggests that decision-support systems should acknowledge the processual nature of affect rather than treating emotion as a single-point influence on outcomes.

\subsection{Towards Adaptive, Affect-Aware Data Informing}
Our results suggest that affect relates to decision-making through complex, non-linear pathways rather than a universal sequence. Notably, each framing engaged a dominant affective domain: Group A, affect-as-information with strong incidental affect (an inward focus on personal discovery, often disconnected from the donation); Group B, affect-as-spotlight with integral affect (attention to the social and institutional context relevant to sharing); and Group C, affect-as-motivation with high variance (deep engagement or disengagement, depending on whether participants could locate their contribution).

Group B's success through social comparison mirrors established findings: Allcott's\cite{allcott_social_2011} evaluation of the Opower program showed that comparing household energy use to similar neighbors reduced consumption by 1.4--3.3\% across 17 experiments. Our study indicates that social comparison can also be operationalized through data visualization in data donation contexts.

These domain-specific effects suggest an opportunity for adaptive systems that dynamically shift framing based on observed user engagement and affective responses. Rather than treating any one framing as universally superior, we envision affect-aware systems that offer multiple lenses or enable lens-shifting. Such a system might detect affective disengagement (as seen in Group C) and shift toward more personally relevant framing, or guide users through a sequence of perspectives: beginning with personally engaging views to establish motivation, then transitioning to social comparison for decision-relevant context. This adaptive approach acknowledges the significant variation in donation rates (37.5\% to 87.5\%) while respecting user autonomy, supporting informed decision-making rather than optimizing for a single outcome.

\section{Limitations and Future Work}
Our study has several methodological limitations, including a relatively small sample size (24 participants, 8 per condition), a controlled research environment that may not reflect real-world donation contexts, 
and a focus on a single data type. Although institutional calendar data can expose personal routines and relationships, it is less sensitive than health, financial, or persistent location data; donation and affective responses may differ for more sensitive data, requiring re-evaluation of the adaptive-framing directions in Section V-D.
With no a priori power analysis, our findings should be read as exploratory rather than confirmatory. Participants were predominantly female, reflecting ID's graduate-only, majority-female student body. Gender was not balanced by design, though prior data-donation research identifies motivational factors such as social duty, rather than demographics, as the primary predictors of donation \cite{skatova_psychology_2019}. Because participants came from the researchers' own institution, institutional culture rather than framing alone may also have shaped willingness to donate.

Future work should test other visualization types and higher-sensitivity data with larger, more diverse samples, examine whether framing effects persist longitudinally, and develop adaptive systems that adjust framing based on detected affective responses. It should also investigate how cultural and demographic differences shape responses to each frame.

\section{Conclusion}
Our study suggests that framing approaches are associated with the affective dimensions of data donation decision-making, with the individual-collective lens (87.5\% donation rate) appearing to best balance personal relevance with social context. The three affective mechanisms (motivation, spotlight, and information) operated differently across framing conditions, with each approach activating a distinct dominant mechanism.

These findings extend affective computing beyond interface-level emotional design to how conceptual framing shapes affective engagement across the data donation decision journey. For practitioners, framing visualizations to balance personal relevance with social context appears most supportive of informed, positive donation decisions. Rather than treating affect as a bias to eliminate, we argue for embracing decision-relevant affective engagement as part of informed consent.

\section*{Ethical Impact Statement}
Data donation is a fraught ethical space with clear potential for abuse, manipulation, and presumptions of false generalizability. Accordingly, considerations relating to privacy, accessibility, and representation were paramount in our research protocol planning and implementation. Our study was conducted with advance approval from Illinois Institute of Technology's Institutional Review Board (approval code IRB-2025-47).

\subsection{Participant Protections}

\subsubsection{Informed Consent}

All participants affirmed their informed consent before accessing and exploring their calendar data with us. The consent process outlined the purpose of the study, the type of data being collected, how it would be used, and that participants would have an opportunity to decide whether to donate their data after the exploration phase. Participants were assured that their participation was entirely voluntary, and that they could leave the experiment at any time, for any reason.

\subsubsection{Data Privacy Protections}

Calendar data contains potentially sensitive information about participants' activities, relationships, and behavioral patterns. To protect the privacy of participants, we were cautious and conscientious in the design of our experiment’s web browser-based experience and data handling processes.

\begin{itemize}[leftmargin=1em]
    \item After accessing data from the Google Calendar API, all calendar data was processed and presented locally during the exploration phase.
    \item Only anonymized data was collected from those who chose to donate, with all human names, physical and virtual locations, and event titles replaced by alphanumeric UUIDs.
    \item All personal identifiers were removed before storage and analysis.
    \item Donated data was stored securely on password-protected hardware, with access restricted to CITI human subject research ethics-certified researchers.
    \item Raw, personally-identifiable recordings such as transcripts and audio files were deleted immediately after coding. 
    \item All anonymized data has been retained, and is available to any interested researcher by contacting the authors.
\end{itemize}

\subsubsection{Coercion}

We carefully designed the study to avoid real or perceived coercion. Participants had complete freedom to decline donation after data exploration, and 37.5\% of participants chose not to donate, demonstrating that participants felt comfortable declining. Because participants were students at the researchers' institution, our recruitment and consent materials stated, in writing and verbally, that donating would not affect their academic standing or institutional relationship. We acknowledge that residual influence may nonetheless remain.

\subsubsection{Accessibility Considerations}

We designed our data exploration interface to be accessible to participants with diverse needs. All visualizations included reduced motion to limit animations, a high-contrast mode, an alternative palette designed for color vision differences, redundant encoding through shape and color, full keyboard navigability, and text alternatives for visual content. We used the Lexend typeface \cite{shaver-troup_lexend_2018} throughout, and our written materials scored 74/100 on the Flesch Reading Ease test. These accommodations support both ethical inclusivity and the validity of affective computing research, as accessibility barriers could confound participants' emotional experiences.

\subsection{Potential Risks and Limitations}

\subsubsection{Affective Manipulation Concerns}

Our findings demonstrate that framing variations influence donation rates. This raises an ethical question about the line between supporting informed decision-making and manipulating it through emotional appeals.
We acknowledge this tension, but given the growing importance of data donation for inclusive and trustworthy AI, we consider these risks acceptable only when framing improves understanding rather than persuades. Our aim is not to maximize donation. The criterion is integral, decision-relevant affect that supports informed choice, not incidental affect that merely lifts the donation rate; a more informed decision to decline is an equally valid outcome.

\subsubsection{Research Experiment Logistical Concerns}

Due to constraints of our institutional environment, we ran our study in an open floor-plan space. We took steps to protect participants' privacy, using mobile partitions and scheduling sessions during non-peak occupancy hours to minimize disruption. Even with these accommodations, some participants may have felt exposed during their sessions.

\subsubsection{Generalizability Limitations}

Our study was conducted with graduate students at a single institution, which limits generalizability to other populations. Cultural variations in attitudes toward privacy, data sharing, and emotional responses to framing may yield different results in other contexts. Our participants did span a globally diverse range of backgrounds, which lends some breadth to the findings, but the single-institution setting remains a constraint. We encourage other researchers to replicate and extend this methodology with larger, more diverse populations across gender, linguistic, sociocultural, ability, and other dimensions.

\subsubsection{Risks of Deception and Manipulation}

Our findings on framing effects could be misused to maximize donation rates through manipulative design. We strongly emphasize that our research aims to improve informed consent practices, not to provide tools for manipulation. The goal should be to support informed decision-making, rather than maximize donation rates.



\section*{Acknowledgment}

The authors are grateful to all participants from the Institute of Design at Illinois Institute of Technology, whose willingness to share their time, data, and reflections made this research possible. We thank Professor Ruth Schmidt for her valuable feedback and expertise in decision-making and behavioral design, which helped strengthen both the theoretical foundation and interpretation of this work. We are also grateful to the anonymous reviewers for their constructive comments.





\bibliographystyle{IEEEtran}
\bibliography{references}




\end{document}


\title{Me Among Us: Affective Framing in Data Donation\\
{\large \textsuperscript{*}Supplementary Material}}

\author{\IEEEauthorblockN{Zeya Chen}
\IEEEauthorblockA{\textit{Institute of Design (ID)} \\
\textit{Illinois Institute of Technology}\\
Chicago, USA\\
zchen103@hawk.illinoistech.edu}
\and
\IEEEauthorblockN{Zach Pino}
\IEEEauthorblockA{\textit{Institute of Design (ID)} \\
\textit{Illinois Institute of Technology}\\
Chicago, USA\\
zach.pino@illinoistech.edu}
}

\maketitle
\thispagestyle{fancy}

This document provides supplementary materials referenced in the main paper, ``Me Among Us: Affective Framing in Data Donation.'' It is not required to follow the paper's argument; it offers additional detail to support reproducibility and assessment. Sections are prefixed ``S,'' tables ``ST,'' and figures ``SF,'' as cited from the main text.

\section{Participant Demographics}
The main paper reports that the three framing groups were broadly comparable in composition. Table~\ref{tab:demo-summary} summarizes participant demographics per framing condition, and Table~\ref{tab:demo-participant} provides participant-level detail. Prior data-donation experience is reported as categorical survey options, and privacy-risk awareness is the group mean on a 1--7 agreement scale; the full items appear with the survey instruments in Section~\ref{sec:instruments}. Together the tables indicate that the groups were comparable in gender, academic level, disciplinary and national background, and prior data-donation experience, supporting the interpretation that the observed donation differences reflect framing rather than group composition.

\begin{table}[!ht]
\caption{Per-condition participant demographics ($N=24$, 8 per group).}
\label{tab:demo-summary}
\begin{center}
\footnotesize
\begin{tabular}{lccc}
\toprule
\textbf{Measure} & \textbf{Group A} & \textbf{Group B} & \textbf{Group C} \\
\midrule
N & 8 & 8 & 8 \\
Gender (M / F) & 3 / 5 & 1 / 7 & 2 / 6 \\
Academic level (Master's / PhD) & 6 / 2 & 7 / 1 & 7 / 1 \\
Age range & 25--40+ & 18--40+ & 18--40+ \\
Prior exp.: none / aware-only & 4 & 5 & 5 \\
Prior exp.: has participated & 3 & 3 & 2 \\
Prior exp.: unsure & 1 & 0 & 1 \\
Privacy-risk awareness, M (1--7) & 3.9 & 3.9 & 2.8 \\
Donation rate & 62.5\% & 87.5\% & 37.5\% \\
\bottomrule
\end{tabular}
\end{center}
\end{table}

\section{Survey Instruments}
\label{sec:instruments}
All items below are reproduced as administered under protocol IRB-2025-47. Three instruments were used: a Demographic and Background Survey, a Pre-Study Survey, and an After-Study Survey.

These instruments provide the measures reported in the main paper: prior data-donation experience (Table~\ref{tab:demo-summary}) corresponds to Demographic Q7; pre-study privacy-risk awareness (Table~\ref{tab:demo-summary}) corresponds to the privacy-risk item of Pre-Study Q5; the pre/post understanding change (Section IV-C) compares the Pre-Study Q5 battery with the After-Study Q5 battery; perceived helpfulness (Section IV-B) corresponds to the ``I know enough to make my decision about whether to donate'' item of After-Study Q5; and post-decision feelings (Section IV-D) correspond to After-Study Q1.

\subsection{Demographic and Background Survey}
\begin{enumerate}
\item \textit{Age:}
\begin{itemize}
\item 18--24
\item 25--29
\item 30--34
\item 35--39
\item 40 or above
\item prefer not to say
\end{itemize}
\item \textit{Gender:}
\begin{itemize}
\item female
\item male
\item non-binary
\item prefer to self-describe
\item prefer not to say
\end{itemize}
\item \textit{Current program at ID.} (open response)
\item \textit{Current semester at ID.} (open response)
\item \textit{Professional/academic background before ID.} (open response)
\item \textit{[Optional] Country/area where you completed your previous degree(s).} (open response)
\item \textit{Prior experience with data donation, sharing, or similar initiatives (e.g., citizen science, crowdsourcing)?}
\begin{itemize}
\item no experience and no prior knowledge
\item no direct experience, but aware of the concept
\item yes, have participated without realizing it was data donation/sharing
\item yes, have knowingly participated once or twice
\item yes, have participated multiple times
\item not sure if my experiences are relevant
\end{itemize}
An optional open response invited a brief description.
\end{enumerate}

\begin{table*}[!t]
\caption{Per-participant demographics. Geographic backgrounds are summarized into broad regions, and donation decisions are omitted, to preserve participant anonymity.}
\label{tab:demo-participant}
\begin{center}
\footnotesize
\begin{tabular}{ccccccc}
\toprule
\textbf{Group} & \textbf{P\_ID} & \textbf{Gender} & \textbf{Age} & \textbf{Program} & \textbf{Academic Background} & \textbf{Geographic Background} \\
\midrule
A & p1 & M & 30--34 & Masters & Design & North America \\
A & p2 & M & 25--29 & Masters & Business & South Asia \\
A & p3 & F & 25--29 & Masters & Business & North America \\
A & p4 & M & 40+ & PhD & Design & North America \\
A & p5 & F & n/a & PhD & Management & North America \\
A & p6 & F & 35--39 & Masters & Psych. / soc. sci. & North America \\
A & p7 & F & 30--34 & Masters & Art & East Asia \\
A & p8 & F & 25--29 & Masters & Design & East Asia \\
\cmidrule(lr){1-7}
B & p9 & F & 18--24 & Masters & Art & East Asia \\
B & p10 & F & 40+ & Masters & Languages / business & Europe \\
B & p11 & M & 25--29 & Masters & Design & South Asia \\
B & p12 & F & 25--29 & PhD & Design & East Asia \\
B & p13 & F & 30--34 & Masters & Social science & North America \\
B & p14 & F & 25--29 & Masters & Design & South Asia \\
B & p15 & F & 25--29 & Masters & Engineering & North America \\
B & p16 & F & 18--24 & Masters & Psychology & North America \\
\cmidrule(lr){1-7}
C & p17 & F & 18--24 & Masters & Architecture & South Asia \\
C & p18 & F & 25--29 & Masters & Design & East Asia \\
C & p19 & M & 25--29 & Masters & Design & North America \\
C & p20 & F & 25--29 & Masters & Law & East Asia \\
C & p21 & F & 25--29 & Masters & Architecture & Europe \\
C & p22 & F & 40+ & PhD & Humanities & North America \\
C & p23 & M & 25--29 & Masters & Engineering / business & North America \\
C & p24 & F & 30--34 & Masters & Architecture & North America \\
\bottomrule
\end{tabular}
\end{center}
\end{table*}

\subsection{Pre-Study Survey}
\begin{enumerate}
\item \textit{How long have you been using Google Calendar?} (not limited to the id.iit.edu account)
\begin{itemize}
\item less than 1 year
\item 1--3 years (since enrolling at ID)
\item 3--5 years
\item more than 5 years
\item other
\end{itemize}
\item \textit{How frequently do you use Google Calendar?}
\begin{itemize}
\item daily
\item several times a week
\item a few times a month
\item rarely
\item never
\end{itemize}
\item \textit{How would you rate your knowledge of Google Calendar functions and features?}
\begin{itemize}
\item lack of knowledge (rarely use it)
\item basic (know only fundamental functions)
\item familiar (comfortable with common features)
\item knowledgeable (confident with most features and functions)
\item expert (confident with most common features and have explored advanced settings/features)
\end{itemize}
\item \textit{How do you commonly access your ID calendar?} (check all that apply)
\begin{itemize}
\item web browser
\item desktop apps (Google Calendar, Apple Calendar, other)
\item mobile apps (Google Calendar, Apple Calendar, other)
\item wearable device
\item integration with other platforms (Zoom, Microsoft Teams, Reclaim.ai, other)
\item others
\end{itemize}
\item \textit{To what degree do you agree with the following statements that describe yourself?} Each item below was rated on a 7-point Likert scale (1: strongly disagree, 2: disagree, 3: somewhat disagree, 4: neither agree nor disagree, 5: somewhat agree, 6: agree, 7: strongly agree).
\begin{itemize}
\item I know who has access to my id.iit.edu calendar data.
\item I understand what types of data and information are included in my id.iit.edu calendar.
\item I understand what types and how much data would be collected if I choose to donate.
\item I understand the purpose and goal of this research.
\item I understand how researchers (data receivers) would process the donated calendar data.
\item I know whether and what ID/IIT administrators can `see' from students' calendars.
\item I understand how donating my data could contribute to the ID community and what collective benefits would be like.
\item I know whether donating my data could bring personal benefits/values beyond helping the community.
\item I understand what potential information exposure and privacy risks I might face if I donate my data.
\end{itemize}
\end{enumerate}

\subsection{After-Study Survey}
\begin{enumerate}
\item \textit{How do you feel about your donation decision?}
\begin{itemize}
\item I feel completely confident and right about my decision
\item I feel generally good and comfortable about my decision
\item I have mixed feelings about my decision
\item I feel somewhat uncomfortable about my decision
\item I feel very uncomfortable about my decision
\item I have no particular feelings about my decision
\item I'm not sure how I feel about my decision
\item other feelings (please specify)
\end{itemize}
An optional open response asked participants to explain why.
\item \textit{During the data exploration session, how impactful was each data visualization?} Each item below was rated from 0 (not impactful at all) through 1--5 (least to most impactful), with n/a for no opinion:
\begin{itemize}
\item time-focused heat map
\item event type/count-focused dot plot
\item connection-focused network map
\item personalized data creation
\end{itemize}
An optional open response asked participants to explain their rating.
\item \textit{Anything missing but you hope to include before donation decision-making or during the data exploration?} (open response)
\item \textit{What could have led you to make different donation decisions?} (open response)
\item \textit{After exploring your data, to what degree do you agree with the following statements that describe yourself?} Each item below was rated on a 7-point Likert scale (1: strongly disagree, 2: disagree, 3: somewhat disagree, 4: neither agree nor disagree, 5: somewhat agree, 6: agree, 7: strongly agree).
\begin{itemize}
\item I know more about the features and functions of Google Calendar.
\item I know who might have access to my data.
\item I understand what and how much information my ID calendar data contains.
\item I understand what types and how much data would be collected for donation.
\item I understand the purpose and goal of this research.
\item I understand how researchers (data receivers) could process the donated calendar data.
\item I know what ID/IIT admins could `see' from students' calendars.
\item I understand how donating my data could contribute to the ID community and what collective benefits are like.
\item I know what personal benefits/values I could gain beyond the collective purpose of donation.
\item I understand what potential information exposure and privacy risks I face if I choose to donate my data.
\item I know enough to make my decision about whether to donate.
\end{itemize}
\item \textit{What's your biggest takeaway from exploring your data? Can you describe specifically what you learned/gained/discovered from this experience?} (open response)
\end{enumerate}

\section{Task Formulations and Think-Aloud Prompts}
Participants were asked to think aloud continuously while exploring each visualization. The opening instruction was: ``As you explore each visualization, please think aloud and say whatever comes to mind. Describe what you notice, what you understand or find confusing, and how you feel about it. There are no right or wrong answers.''

When a participant fell silent, the facilitator used non-directive probes that did not lead toward donating or declining:
\begin{itemize}
\item ``What are you noticing here?'' / ``What do you make of this?''
\item ``How do you feel about what you are seeing?'' / ``Why do you feel that way?''
\item ``Does this change how you think about donating your calendar data, and how?''
\item ``What, if anything, makes you more or less comfortable about donating?''
\end{itemize}
The affect-focused probes (``how do you feel'' / ``why'') elicited the in-the-moment affective expressions coded in the affect-as-information analysis (Section IV-D).

\section{Affective Coding for the Affect-as-Information Analysis}
The codebook below supports the affect-as-information analysis (Section IV-D); the affect-as-motivation and affect-as-spotlight results rest on quantitative measures (Section~\ref{sec:instruments}) and were not thematically coded. Following reflexive thematic analysis, a single analyst interpretatively coded each participant's affective expressions, drawn from think-aloud comments during exploration (in-the-moment affect) and open-ended post-study responses (retrospective affect); post-decision feeling ratings are from after-study Q1. Coding began with open coding of emotion expressions using the codes below; the resulting affective patterns, and their association with donation decisions and framing condition, are reported in Section IV-D. We report these findings as exploratory and illustrative, so inter-rater reliability is not used as a quality criterion.

\subsection{Affect (emotion) codes}
\begin{itemize}
\item INTEREST: curiosity or interest in the visualizations or interaction.
\item COMFORT: comfort or confidence with the data being shown or shared.
\item SURPRISE: surprise or delight at what the data revealed.
\item CONFUSION: confusion or difficulty understanding the data or visualizations.
\item WORRY: worry or concern, for example about privacy or information exposure.
\item SKEPTICISM: doubt or skepticism, for example about data accuracy or security.
\end{itemize}

\subsection{Integral vs.\ incidental marking}
Each affective expression was marked \textit{integral} when it concerned a topic relevant to the donation decision (e.g., privacy and information exposure, data use and access, or collective benefit) and \textit{incidental} when it did not (e.g., aesthetic delight or personal-memory recall). Group A's affect was predominantly incidental and Group B's predominantly integral.

\subsection{Dominant reading by condition}
Affective expressions also differed in interpretive stance: Group A was largely self-focused, Group B comparative (self relative to others), and Group C collective, with some confusion about whose data was shown. These readings align with each condition's dominant mechanism (Group A: information/incidental; Group B: spotlight/integral; Group C: motivation, high variance).

\subsection{Scope}
Initial coding also captured non-affective dimensions (e.g., stated donation reasons, perceived gains and losses, and calendar-usage or data-governance topics). These provided context but are not part of the affective analysis reported here.

\section{Framing Text and Visualizations by Condition}
\label{sec:framing}
\begin{figure*}[!t]
\centerline{\includegraphics[width=\textwidth]{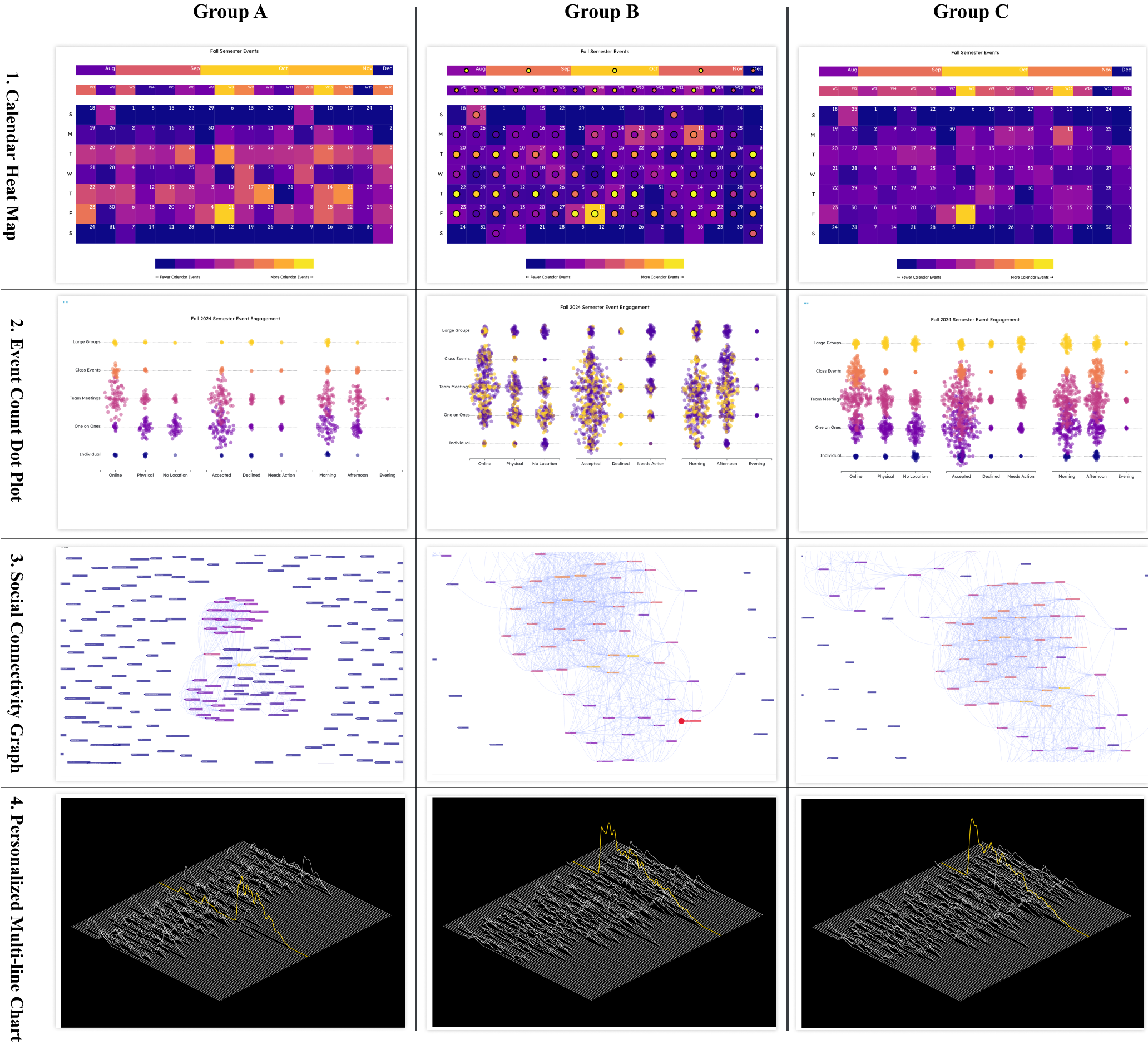}}
\caption{The four visualization archetypes (rows) as rendered in each framing condition (columns: Groups A, B, and C). All twelve views are generated from the same example participant's donated calendar data; each condition saw the same archetypes, differing only in data comparison, tooltip text, and on-screen instructions.}
\label{fig:vis-all}
\end{figure*}

All conditions viewed the same four interactive visualization archetypes (previewed in Fig.~1 of the main paper); only the data comparison, tooltip text, and on-screen instructions differed by condition. Fig.~\ref{fig:vis-all} shows all four archetypes as each condition rendered them, generated from the same example participant's donated calendar data, so any visible differences across conditions reflect the framing design rather than the underlying data. Figs.~\ref{fig:vis-heatmap} through \ref{fig:vis-multiline} then present one annotated close-up per archetype, with annotations marking the per-condition differences. Names and email addresses shown in the tool are anonymized.

\subsection{Platform}
\begin{figure*}[!t]
\centerline{\includegraphics[width=\textwidth]{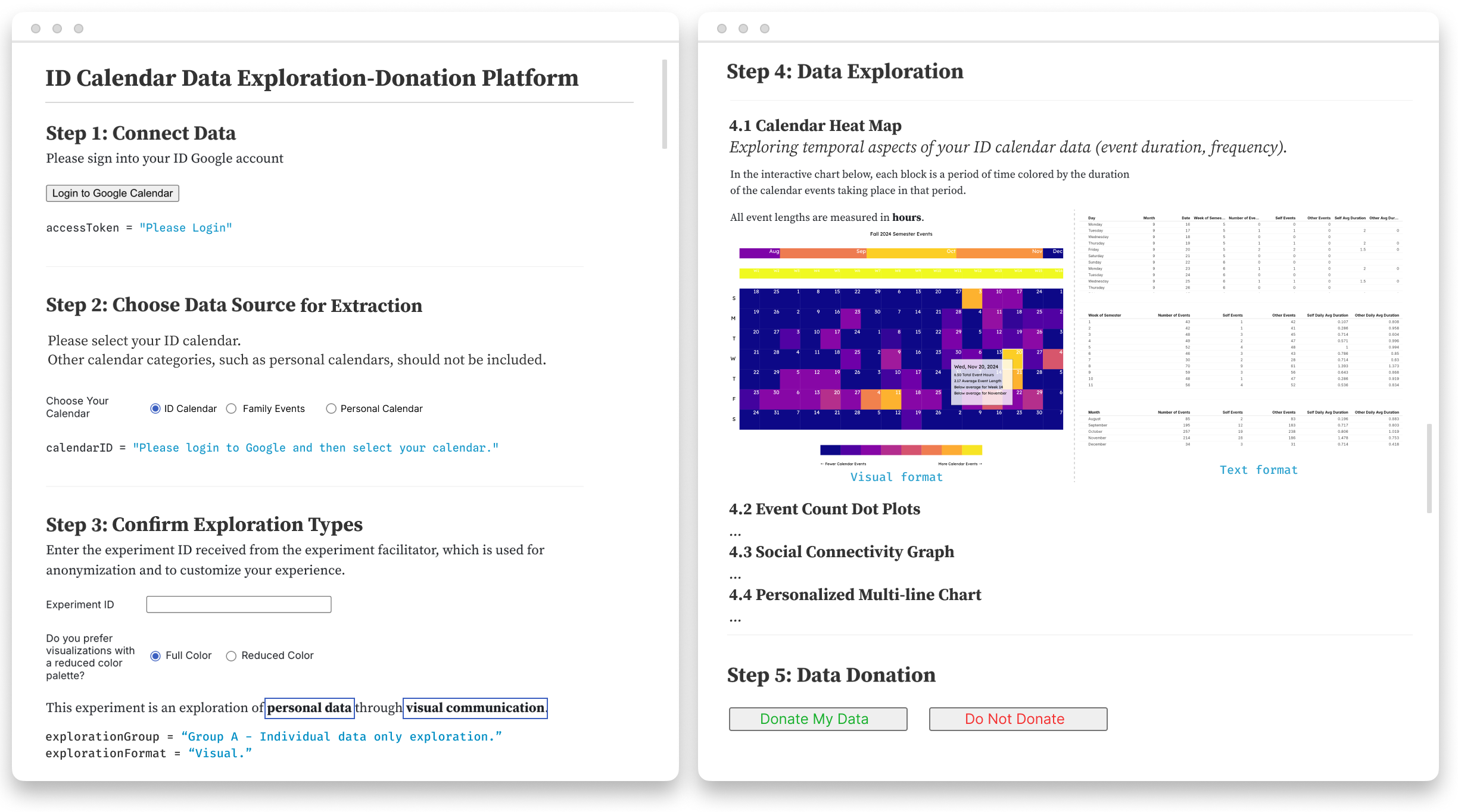}}
\caption{The exploration and donation platform. Left: Steps 1 to 3, where participants sign into their institutional Google account, select the calendar for extraction, and confirm their assigned exploration group. Right: Steps 4 and 5, where participants explore the four visualization archetypes before making the donation decision.}
\label{fig:platform}
\end{figure*}

Fig.~\ref{fig:platform} shows the privacy-preserving web platform we designed and developed for data exploration and donation. Participants signed in with their institutional Google Calendar credentials to access their data, which was processed locally in the browser to generate their assigned condition (A, B, or C); no data left the device unless the participant chose to donate. Participants then scrolled through the four visualization archetypes in sequence, with the raw data (table and JSON) available for optional inspection, before making the donation decision. To protect privacy, researchers could not view the participant's screen, and after deciding, the participant logged out and verified that the platform retained no data.

\subsection{Visualization Archetypes}
\begin{figure*}[!t]
\centerline{\includegraphics[width=0.8\textwidth]{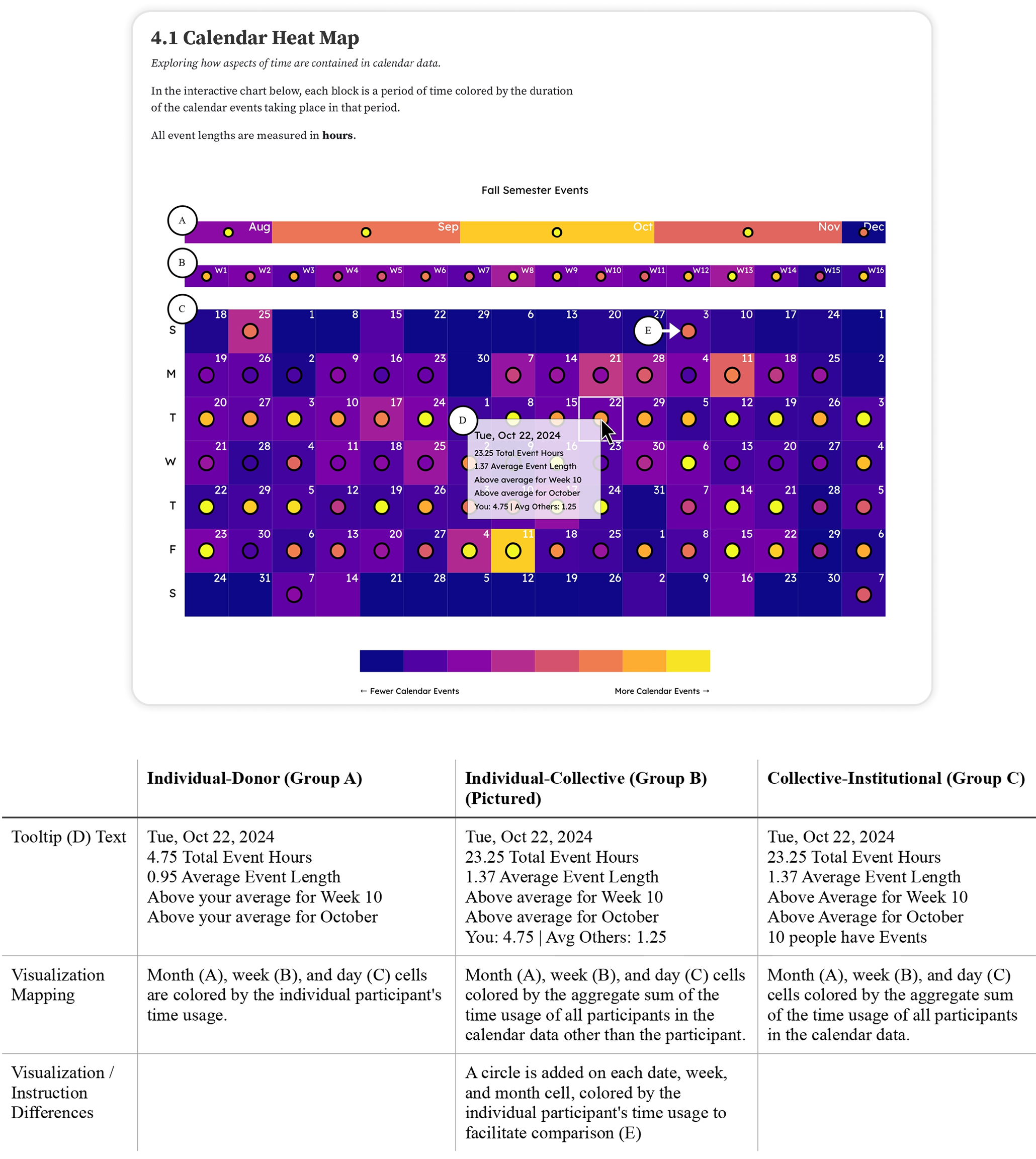}}
\caption{Calendar Heat Map (Group B pictured). Month, week, and day cells are colored by the duration of calendar events in each period, letting participants explore how their time is distributed. Annotations mark the per-condition differences.}
\label{fig:vis-heatmap}
\end{figure*}

\begin{figure*}[!t]
\centerline{\includegraphics[width=0.8\textwidth]{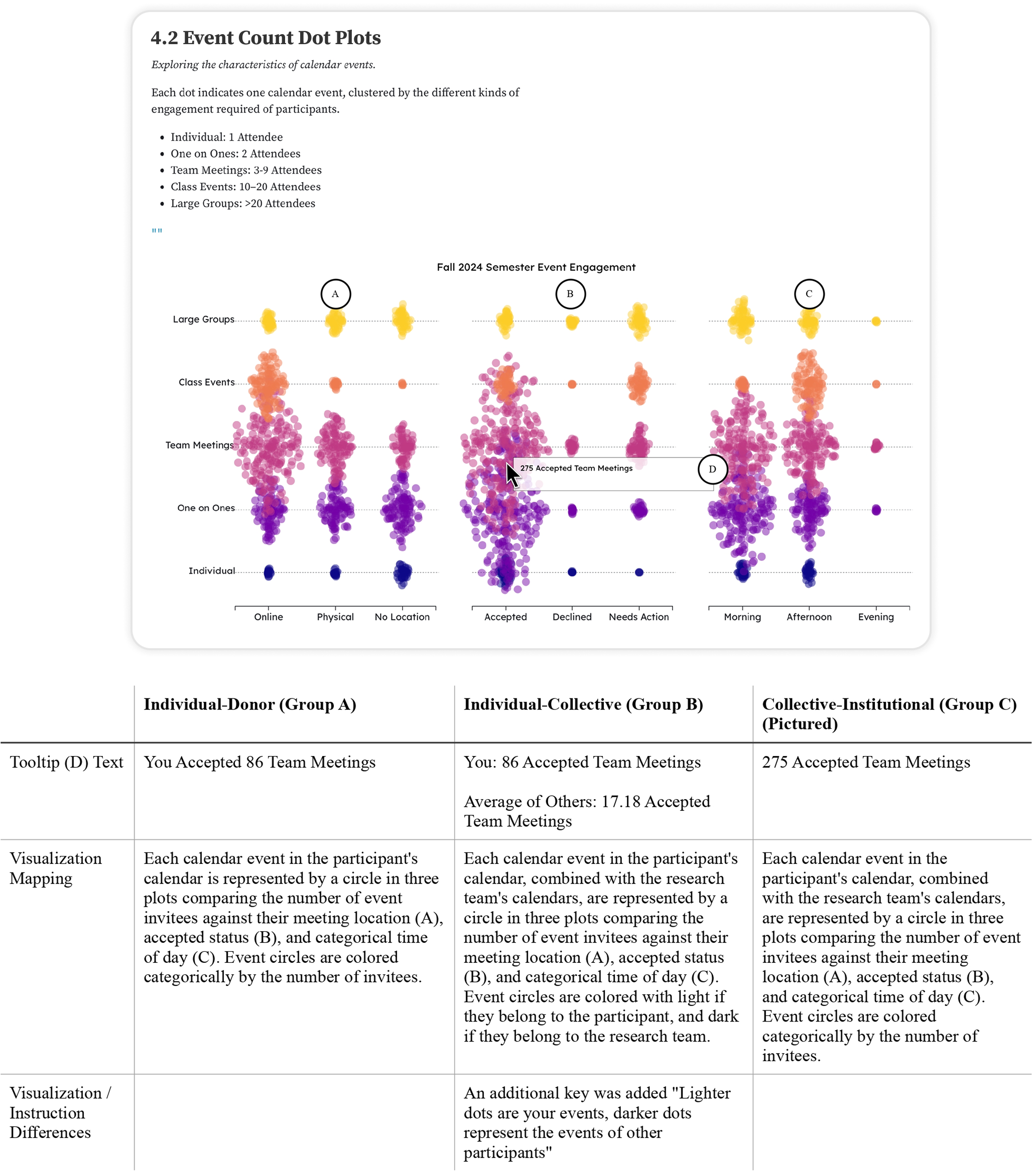}}
\caption{Event Count Dot Plots (Group C pictured). Each dot is one calendar event, clustered by engagement type; three linked plots compare the number of invitees against meeting location, accepted status, and time of day. Annotations mark the per-condition differences.}
\label{fig:vis-dotplot}
\end{figure*}

\begin{figure*}[!t]
\centerline{\includegraphics[width=0.8\textwidth]{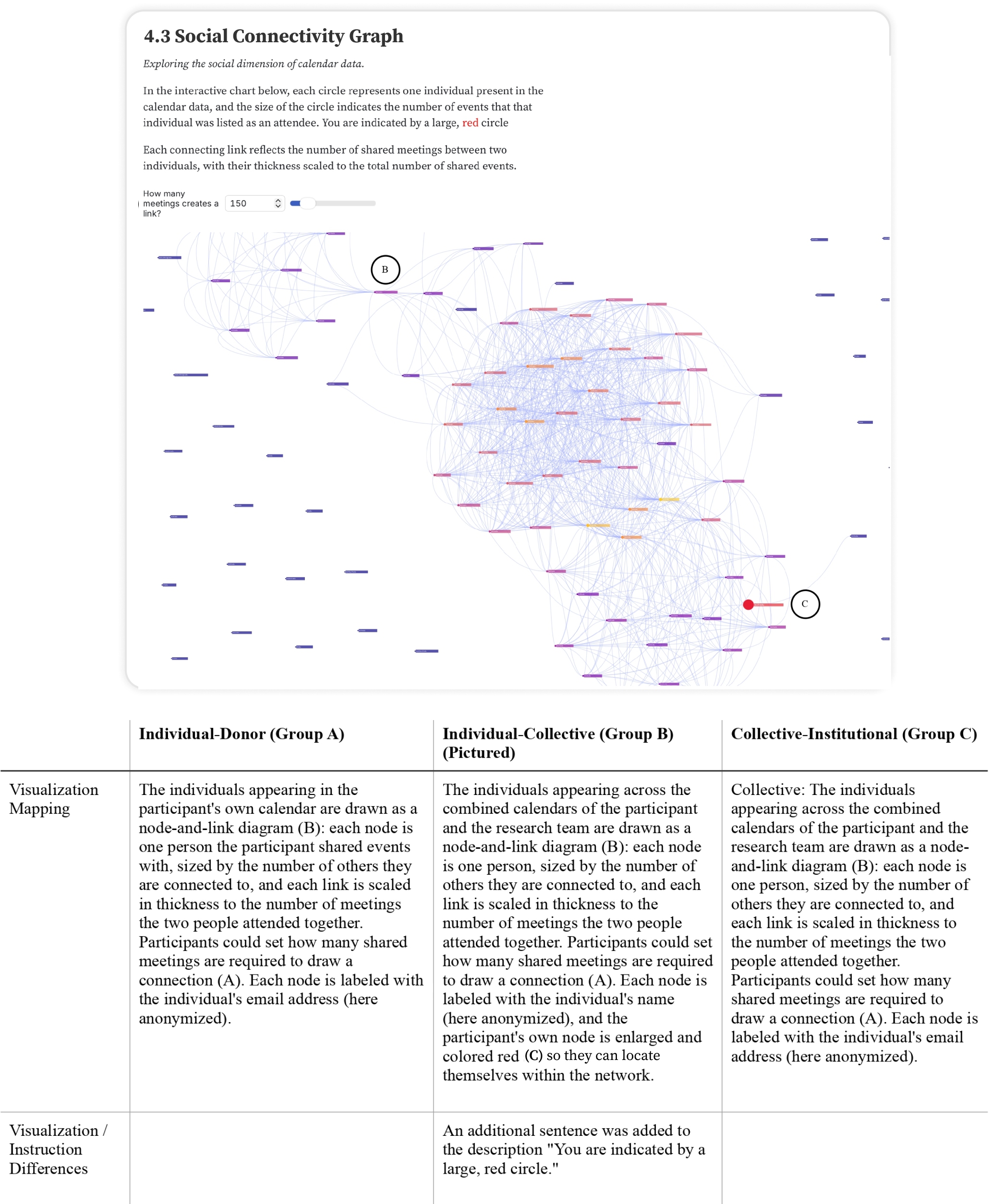}}
\caption{Social Connectivity Graph (Group B pictured). Each node is a person, sized by how many others they connect to; each link's thickness scales with the number of shared meetings. Annotations mark the per-condition differences.}
\label{fig:vis-network}
\end{figure*}

\begin{figure*}[!t]
\centerline{\includegraphics[width=0.8\textwidth]{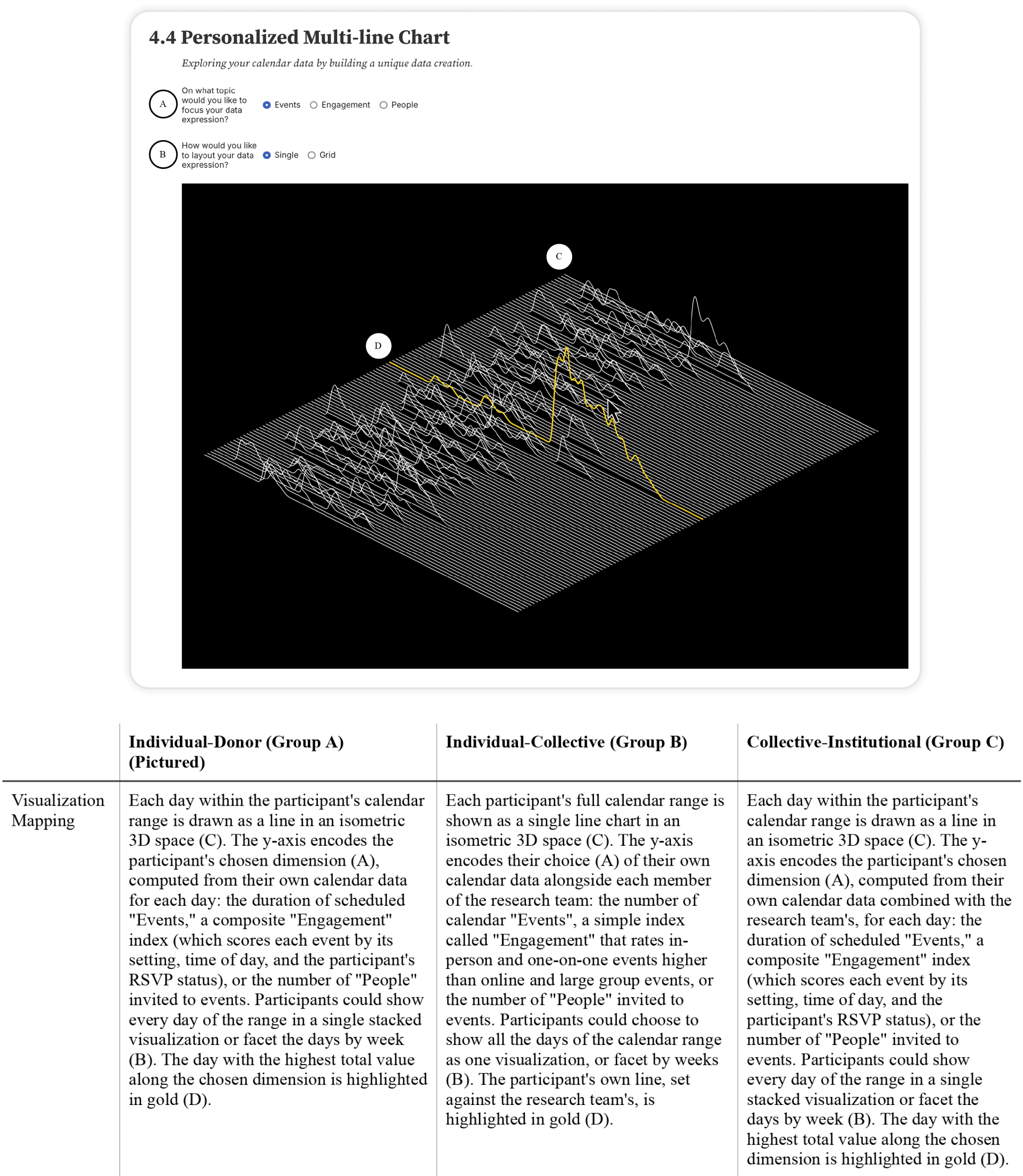}}
\caption{Personalized Multi-line Chart (Group A pictured). Each day in the participant's calendar range is drawn as a line in an isometric 3D space, with a y-axis the participant chooses: event duration, an engagement index, or the number of people invited. Annotations mark the per-condition differences.}
\label{fig:vis-multiline}
\end{figure*}

Figs.~\ref{fig:vis-heatmap} through \ref{fig:vis-multiline} present one annotated close-up per archetype; each caption describes the archetype, and the annotations mark the per-condition differences.

